\documentclass[journal]{ieeetran}
\usepackage{cite}
\usepackage{amsmath,amssymb,amsfonts}
\usepackage{algorithmic}
\usepackage{graphicx}
\usepackage{textcomp}
\usepackage{mathrsfs}

\usepackage{bbding}
\usepackage{pifont}
\usepackage{wasysym}
\usepackage{amssymb}
\newcommand{\xmark}{\ding{55}}%
\usepackage{makecell}
\usepackage{multirow}
\usepackage{graphicx}
\usepackage{tikz}
\usepackage{tabu}
\usepackage{pgfplots}
\usepackage{fancyhdr}
\usepackage{hyperref}

\usepackage{hyphenat}
\definecolor{LightCyan}{rgb}{0.88,1,1}
\usepackage[first=0,last=9]{lcg}

\usepackage{booktabs}
\usepackage[flushleft]{threeparttable}
\usepackage{epstopdf}
\usepackage{mathtools}
\usepackage{amsmath}
\graphicspath{{figures/}}
\usepackage[labelfont=bf]{caption}
\usepackage{color,soul}
\usepackage{bm}
\usepackage{float}
\usepackage{xcolor}

\def\BibTeX{{\rm B\kern-.05em{\sc i\kern-.025em b}\kern-.08em
    T\kern-.1667em\lower.7ex\hbox{E}\kern-.125emX}}

\title{ASL to PET Translation by a Semi-supervised Residual-based Attention-guided Convolutional Neural Network}

\usepackage{tikz}
\usepackage{amsmath}
\usepackage{hyperref}
\usepackage{array}
\usepackage{wrapfig}
\usepackage{multirow}
\usepackage{xcolor}
\usepackage{subcaption}
\usepackage{authblk}
\usepackage{pdfpages}
\usepackage{xparse}
\usepackage{tabu}
\usepackage{wrapfig}
\usepackage{xcolor}
\usepackage{soul}
\usepackage{pgfplots}
\usepackage{amssymb}
\usepackage{flexisym}

\usepackage{amsmath}
\usepackage{mathtools}
\usepackage{xcolor}
\usepackage{lipsum} 
\usepackage[colorinlistoftodos]{todonotes}
\usepackage{pdfcomment}

\usepackage[T1]{fontenc}
\DeclareFontFamily{T1}{calligra}{}
\DeclareFontShape{T1}{calligra}{m}{n}{<->s*[1.44]callig15}{}
\DeclareMathAlphabet\mathcalligra   {T1}{calligra} {m} {n}
\DeclareMathAlphabet\mathzapf       {T1}{pzc} {mb} {it}

\makeatletter
\newcommand*{\rom}[1]{\expandafter\@slowromancap\romannumeral #1@}
\makeatother

\author[1]{Sahar~Yousefi}
\author[1]{Hessam~Sokooti}
\author[1]{Wouter~M.~Teeuwisse}
\author[2]{Dennis~F.R.~Heijtel}
\author[3]{Aart~J.~Nederveen}
\author[1,4]{Marius~Staring}
\author[1]{Matthias~J.P.~van Osch}

\affil[1]{Department of Radiology, Leiden University Medical Center, Leiden, The Netherlands}
\affil[2]{Philips, Amsterdam, The Netherlands}
\affil[3]{Department of Radiology and Nuclear Medicine, Amsterdam University Medical Center, Amsterdam, The Netherlands}
\affil[4]{Delft University of Technology, Delft, The Netherlands}

\affil[ ]{\textit {s.yousefi.radi@lumc.nl}}

\begin{document}
\maketitle

\begin{abstract}
Positron Emission Tomography (PET) is an imaging method that can assess physiological function rather than structural disturbances by measuring cerebral perfusion or glucose consumption. However, this imaging technique relies on injection of radioactive tracers and is expensive. On the contrary, Arterial Spin Labeling (ASL) MRI is a non-invasive, non-radioactive, and relatively cheap imaging technique for brain hemodynamic measurements, which allows quantification to some extent. In this paper we propose a convolutional neural network (CNN) based model for translating ASL to PET images, which could benefit patients as well as the healthcare system in terms of  expenses and adverse side effects. However, acquiring a sufficient number of paired ASL-PET scans for training a CNN is prohibitive for many reasons. To tackle this problem, we present a new semi-supervised multitask CNN which is trained on both paired data, i.e. ASL and PET scans, and unpaired data, i.e. only ASL scans, which alleviates the problem of training a network on limited paired data. 
Moreover, we present a new residual-based-attention guided mechanism to improve the contextual features during the training process. Also, we show that incorporating T1-weighted scans as an input, due to its high resolution and availability of anatomical information, improves the results. 
We performed a two-stage evaluation based on quantitative image metrics by conducting a 7-fold cross validation followed by a double-blind observer study. The proposed network achieved structural similarity index measure (SSIM), mean squared error (MSE) and peak signal-to-noise ratio (PSNR) values of $0.85\pm0.08$, $0.01\pm0.01$, and $21.8\pm4.5$ respectively, for translating from 2D ASL and T1-weighted images to PET data. The proposed model is publicly available via \url{https://github.com/yousefis/ASL2PET}.


\end{abstract}
\section{Introduction}

Positron emission tomography (PET) is a diagnostic imaging modality which allows assessment of brain hemodynamic parameters, including cerebral blood flow (CBF), cerebral blood volume, and glucose uptake \cite{robert1999positron}. MRI on the other hand provides the ability to measure both brain structure and physiology at the same time, e.g. T1-weighted (T1w) MRI provides brain structure and Arterial Spin Labeling (ASL) sensitizes the MR-signal to perfusion. However, different from PET, which is radioactive and relatively expensive, MRI with ASL is non-invasive, non-radioactive, and relatively cheap. Although ASL has been shown to provide similar measurements as PET, the latter is still considered the gold standard. Therefore, we present a convolutional neural network (CNN) for synthesising PET scans from ASL scans. However, access to PET is scarce, due to the mentioned reasons \cite{jensen2012oxygen}, which is an obstacle for collecting a sufficiently large data set for training a fully supervised CNN. To tackle this problem, we leverage a semi-supervised learning approach for training the proposed network. Semi-supervised learning has been attracting attention in training CNNs due to its applicability in problems in which access to paired data, i.e. PET-ASL scans, is limited while access to unpaired data, i.e. single ASL scans, is more commonly available \cite{zhu2009introduction}. Finally, a crucial step in medical artificial intelligence is validation of the performance when applied to pathological situations. As a surrogate test, we propose the application to a resting and activated condition (either visual or motor activation), since neuronal activation is known to elicit a local increase in blood flow. By testing the network on data in which the normal blood flow pattern is disturbed, the performance of the network under new conditions can be validated to some extent.

Recently, semi-supervised methods have been receiving attention in computer-aided medical applications \cite{cheplygina2019not}, such as segmentation \cite{khosravan2018semi,chen2019multi,terzopoulos2019semi}, classification \cite{imran2020partly}, and regression \cite{zhang2012multi}. In this work we propose a new semi-supervised CNN for ASL and T1w images to PET translation. The CNN-based image-to-image translation approaches can be divided into two groups: encoder-decoder networks and generative adversarial networks (GANs) \cite{kaji2019overview}. 


Yoo \textit{et al.} used two pre-trained auto-encoders and decoder pairs for each source and target image domain with an application to natural images translation \cite{yoo2019image}. 
Han proposed an encoder-decoder architecture \cite{han2017mr} to generate a synthetic CT from brain MR images. Xian \textit{et al.} synthesized CT images from T1w MR images utilizing a deep embedding convolutional neural network. There was no downsampling block in this architecture and the proposed embedding block in this study is similar to deep supervision \cite{lee2015deeply}. In a similar application, Dinkla \textit{et al.} used a dilated convolutional neural network \cite{dinkla2018mr}.

Recently, many GAN-based models for MRI to CT translation have been proposed \cite{kearney2020attention,jin2019deep,kaiser2019mri,wolterink2017mr}. Armanious et al. proposed a fully supervised GAN-based network for medical image to image translation including PET, MRI and CT \cite{armanious2020medgan}. In \cite{armanious2019unsupervised} an unsupervised cycle GAN network \cite{zhu2017unpaired} for PET to CT translation has been proposed. Ben-Cohen et al. \cite{ben2019cross} proposed a GAN network for CT to PET translation. Yang et al. Jin et al. proposed a GAN model for CT to MRI translation \cite{jin2019dc2anet}. Nguyen et al. proposed a semi-supervised adversarial cycleGAN for translating between two MRI neuroimaging modalities \cite{nguyen2020semi}. In \cite{yang2020mri} a GAN network for multi-contrast MRI scans translation, including T1, T2 and PD-weighted, has been proposed. Jung et al. proposed a conditional GAN (cGAN) for anatomical MRI to PET translation \cite{jung2018inferring}. In that study, no physiological information has been considered. In \cite{dar2019image} a multi-contrast MRI to MRI translation method based on a cGAN has been proposed. 
Chen et al. proposed a fully-supervised encoder-decoder based model for predicting cerebrovascular reserve using combined multi-contrast information from baseline PET and MRI \cite{chen2020predicting}. 
In \cite{hu2020brain} a bidirectional GAN was utilized in order to synthesize a PET scan from an MR brain image. The aim was to reflect the diverse brain attributes of different subjects.


This study proposes a multi-task CNN for constructing PET scans in a semi-supervised fashion by MRI scans composed of ASL and T1w scans \cite{van2018advances,yousefi2019fast,yousefi20103d,yousefi2012brain}. ASL provides physiological information in low resolution while T1w images provide anatomical information in a high resolution format. The network is composed of one encoder and two decoders for constructing the PET and regenerating the input ASL scan. In order to encourage the network to extract relevant physiological information, skip connections are only used between the encoder and PET decoder, for retrieving the resolution of the PET scans. The training is performed using a two-step procedure, where we first train the network on the whole scans, and subsequently train on unlearnt regions. We leverage a spatial attention gate for identification and focusing of the network to these unlearnt regions. Although PET and ASL provide similar measurements, they are different modalities and look differently specially in spatial resolution and signal-to-noise ratio (SNR)  (ASL has a higher resolution and SNR). Therefore we employ channel attention gates at the skip connections to aid the network in selectively disentangle the PET relevant features \cite{yousefi2020esophageal}. 
The contributions of this study are as below:
\begin{itemize}
  \renewcommand{\labelitemi}{$\bullet$}
    \item To the best of our knowledge, this is the first semi-supervised multitask CNN approach for medical image translation and specifically for MRI to PET translation.
    \item We propose a new training strategy called residual-based attention-guided step to encourage the network to selectively improve the local physiological context during both training and inference processes.
    \item We propose disentanglement attention gates through the skip connections in order to guide the network to selectively disentangle the features in the proposed multitask CNN properly during both the training and inference processes. 
    \item We perform validation not only on resting state perfusion data, but also on regionally activated data to mimick pathology-induced local perfusion changes.
\end{itemize}

\section{Proposed method}
\subsection{Problem}

\begin{figure*}[hbt!]
  \centering
    \includegraphics[width=17cm,clip]{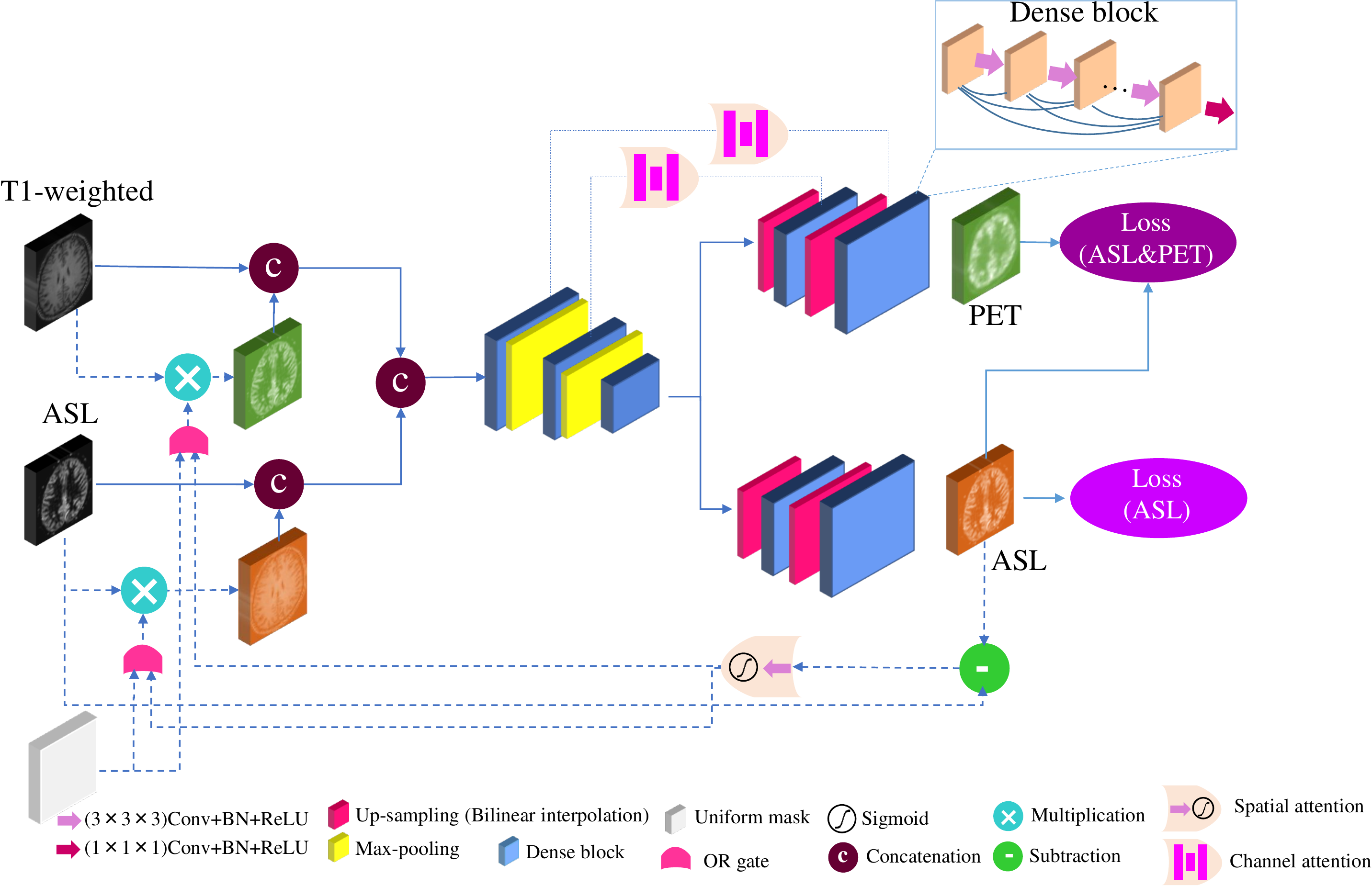}
  \caption{The architecture of the proposed method. }   
  \label{fig:cnn} 
\end{figure*}
In this study, the semi-supervised multitask CNN-based MRI (ASL together with T1w) to PET translation is defined as follows. We have $M$ ASL and T1w scans that are unpaired with a ground truth PET scan:
\begin{equation}
    \left(A^{u},T^{u}\right) = \{(a^{u}_1,t^{u}_1),\ldots,(a^{u}_M,t^{u}_M)\},
\end{equation}
in which $a^{u}_i\in{A}$ and $t^{u}_i\in{T}$ are the ASL and T1w scans of the $i$-th subject. We also have a set of $N$ images that are paired with a PET:
\begin{equation}
    \left(A^{p},T^{p},P^{p}\right) = \{(a^{p}_1,t^{p}_1,p^{p}_1),\ldots,(a^{p}_N,t^{p}_N,p^{p}_N)\},
\end{equation}
in which $a^{p}_i \in {A}^{m\times n}$, $t^{p}_i\in {T}^{m\times n}$ and $p^{p}_i \in {P}^{m^\times n}$ are the ASL, T1w and  PET scans of the $i$-th subject, respectively. Our goal is designing a CNN, ${N}$, as 
\begin{equation}
    p_{j} =  {N}\{(a_{j},t_{j})\},
\end{equation}
which receives unseen $a_{j}$ and $t_{j}$ as the inputs and generates $p_{j}$ as the output, in which $j$ denotes the $j$-th subject.

As described before, different from T1w scans, ASL has low resolution. However, ASL contains the physiological information while T1 provides anatomical information. Therefore, in this paper we propose a multitask CNN-based approach for generating PET scans from both T1 and ASL scans. 

\subsection{Proposed Network Architecture}\label{sec:method}
In this section the proposed network is explained. Figure \ref{fig:cnn} shows the proposed 2D CNN. 

We designed the network with a shared encoder and two PET and ASL branches composed of two separate decoders for generating the PET and ASL scans, respectively. The ASL decoder is utilized to boost the encoder to extract the physiological context as much as possible. Therefore the network is not equipped with skip connections between the encoder and ASL decoder to prevent the ASL decoder from directly obtaining the feature maps from the encoder. The proposed multitask CNN goal can be described as follows:
\begin{align}
    p_i &= {D}_{P}\left({E}\left(a_i,t_i\right)\right),\\
    a_i &= {D}_{A}\left({E}\left(a_i,t_i\right)\right)
\end{align}
in which $ {D}_{P} $ and ${D}_{A}$ are the PET and ASL construction decoders and ${E}$ is the shared encoder, respectively. $ {D}_{P} $ is trained with paired data and ${D}_{A}$ and ${E}$ are trained by both paired and unpaired data.

The network is composed of dense blocks decorated in a Unet shape with a configuration of $[1, 3, 5, 3, 1]$ in which the five numbers denote the number of $(3\times3\times3)$ convolutional layer, batch normalization (BN), and rectified linear unit (ReLU) layers successively in five dense blocks at different levels of the proposed CNN similar to \cite{yousefi2018esophageal, yousefi2020esophageal}. 

Although the proposed CNN is extracting the physiological information from the ASL and anatomical information from the T1w scan, ASL scans have a higher signal to noise ratio compare to PET. Therefore, PET and ASL look slightly different. In order to persuade the network to extract the useful features for synthesizing the PET scans, disentanglement attention gates have been leveraged in the skip connections between ${E}$ and ${D}_{P}$, while there are no skip connections between ${E}$ and ${D}_{A}$. The non-existence of the skip connections between the encoder and the ASL decoder prevents the network from copying the features. This pushes the encoder to extract physiological relevant features independently.

\subsection{Training strategy}
We propose a new alternating training process for each training sample. For every even iteration of the network training we perform coarse training using the T1w and ASL scans as input, to reconstruct the output scans. The coarse details of the outputs, i.e. the ASL and PET scans for the paired data and only the ASL scan for the unpaired data, is generated. For every odd iteration we perform fine-grained training, which is performed to persuade the network to improve the fine details of the outputs. For this goal we propose a residual attention-guided mechanism, which creates an attention map by subtracting the reconstructed ASL from the input ASL and feeding this into a spatial attention gate. Then, the input ASL and T1w scans are multiplied by the probabilistic attention map and fed to the network along with the original input scans. This step encourages the network to concentrate on the regions which may cause a high loss value. Note that in the coarse training step, since the network focuses on constructing the whole scan, the attention mask is a uniform mask with a value of 1 everywhere (see Figure \ref{fig:cnn}). In the fine-grained training step, the uniform mask is replaced by the generated attention map. This is shown by an 'or' operation in Figure \ref{fig:cnn}, where the residual attention-guided mechanism is shown by the dashed arrows. The two-step training process for the input patches of $\{a_i, t_i\}$ can be formulated as follows: 
\begin{itemize}
    \item Training at iteration $j-1$ with  $x_{j-1}=\{[a_i,a_i\times M_{u}], [t_i,t_i\times M_{u}]\}$:
    \begin{align}
    p_{ij} &={D}_{{P}}\left({E}(x_{j-1})\right),\\
    \{a_{i,j},M_{i,j}\} &= {D}_{{A}}\left({E}(x_{j-1})\right),
    \end{align}
    in which $M_{u}$ is the uniform mask, $M_{i,j}$ is the residual-attention mask for the $i-$th input at time $j$, and $[\cdot]$ denotes concatenation.
    
    \item  Training at iteration $j$ with  $x_j=\{[a_{i},a_{i}\times M_{ij}],[t_{i},t_{i}\times M_{ij}]\}$:
    \begin{align}
    p_{i,j+1} &={D}_{{P}}\left({E}_{j}(x_j)\right),\\
    \{a_{i,j+1},M_{i,j+1}\}&=\{{D}_{{{P}}}\left({E}(x_j)\right), {D}_{{A}}\left({E}(x_j)\right)\}.
    \end{align}
\end{itemize}

\subsection{Loss function}\label{sec:loss_func}
Two loss functions for training the network are deployed, one for the data paired with a PET and one for the data unpaired with a PET. For the paired data the loss is computed on the sum of the discrepancy between the reconstructed ASL and PET scans with respect to the ground truth ASL and PET scans, while for the unpaired data this loss is computed only on the ASL. As the efficiency of structural similarity index measure (SSIM) in medical image reconstruction has been discussed and proved in the literature of image reconstruction \cite{yousefi2019fast,pezzotti2020adaptive}, we adopt this measure for calculating the discrepancy between the scans. For the paired data the two discrepancies are weighted equally.

\section{Materials and implementation}\label{sec:experiments}
\subsection{Datasets}\label{sec:training}
\begin{table*}[tb!]
\caption{Details of the datasets}
\begin{center}
\centering
\begin{tabular}{cccccccc}
    \hline
    \multirow{2}{*}{Dataset}& \multicolumn{3}{c}{Scans}&
    \multirow{2}{*}{imaging schedule}&
    \multirow{2}{*}{$\#$ of subjects}& \multirow{2}{*}{Age}&
    \multirow{2}{*}{Scanner}
    \\
    &T1&ASL&PET&&
    \\
    \hline
    \multirow{2}{*}{AMC}& \multirow{2}{*}{\checkmark}& \multirow{2}{*}{\checkmark}& \multirow{2}{*}{\checkmark}&
     \multirow{2}{*}{\begin{tabular}{@{}ll@{}}
                    Session $\#$1 &
                  Session $\#$2\\
                  ASL 2x, PET & ASL, PET\\
                 \end{tabular}}
     &
    \multirow{2}{*}{16 (9 male)}&
    \multirow{2}{*}{20$-$24} & 
    Philips 3 Tesla Intera system \\
    &&&&&&&Philips Gemini TF-64 PET/CT system\\
    \hline
    \multirow{2}{*}{LUMC}&
    \multirow{2}{*}{\checkmark}& 
    \multirow{2}{*}{\checkmark}& 
    \multirow{2}{*}{\xmark}&
    \multirow{2}{*}{\begin{tabular}{@{}ccc@{}}
                    Rest&
                  Visual cortex&
                  Motor cortex\\
                  ASL& 
                  ASL 2x&
                  ASL\\
                 \end{tabular}
                 }
                 &
     \multirow{2}{*}{27 (13 male)}&
     \multirow{2}{*}{19$-$35}&
     \multirow{2}{*}{Philips 3 Tesla Intera system}
     \\\\
    \hline
\end{tabular}
\end{center}
\label{tab:datasets}
\end{table*}

This study includes two datasets: 
\begin{itemize}
  \renewcommand{\labelitemi}{$\bullet$}
    \item AMC dataset \cite{heijtel2014accuracy}: this dataset includes O15-H2O PET and ASL scans of 16 distinct subjects, acquired under baseline (normocapnia, three repeated measurements) and hypercapnia (two repetitions) conditions over two study days. Hypercapnia refers to a condition in which the carbon dioxide (CO$_2$) level in the blood is elevated by breathing air with a higher content of CO$_2$. On a first study day, two normacapnia and one hypercapnia scan were performed in a single imaging session, separated by 20 to 30 minutes. PET and MRI were done in separate sessions, since no MR-PET scanner was available. On a second study day after approximately 28 days, one normocapnia and one hypercapnia scan were made, both on PET and MRI, to assess inter-session reproducibility. We refer to the original article for acquisition and study details \cite{heijtel2014accuracy}. Additionally, in each MRI session a T1w image was acquired. 
    \item LUMC dataset: this dataset includes ASL and T1w scans of 27 subjects under a resting state condition and under three neuronal activation states (visual and motor cortex stimuli), caused by respectively watching a Tom $\&$ Jerry cartoon (visual cortex), watching a 8 Hz flickering checkerboard (visual cortex), and by finger tapping (motor cortex). 
    
\end{itemize}
Table \ref{tab:datasets} tabulates the details of the datasets. 

\subsection{Training details}\label{sec:training}
The network has been developed in Google’s Tensorflow and the experiments have been performed on a NVIDIA Quadro RTX6000 with 24 GB of GPU memory.


For method validation, 2 out of 16 subjects of the AMC dataset and 14/27 of the LUMC data were used for test set, while the remainder (4 out of 6 for AMC and 23 out of 27 for LUMC) was used as an independent model optimization set. On the data used for network optimization a 7-fold cross-validation was conducted. In each fold, this split was divided randomly into 12 (AMC) and 23 (LUMC) subjects for training, and 2 (AMC) and 0 (LUMC) subjects for validation. 
Since the LUMC dataset does not include PET scans and the evaluation measures can not be calculated, there is no LUMC subject in the validation set. Randomization was performed such that no overlap existed between the validation sets of the different folds.

To manage the utilization of the GPU and CPUs during the training process, a multi-threaded daemon process extracts and queues the slices from each of the modalities. 
We used a batch size of 10. The networks were trained for about 100k iterations which implies 50k for fine and 50k for coarse training steps. The batch of samples are extracted either from the AMC or LUMC dataset for training both branches (ASL and PET), or only the ASL branch respectively. For every 5 successive iterations, the training was performed for one of the AMC or LUMC dataset. 



\section{Experiments and results} \label{hd:Evaluation}
In this section the quantitative and qualitative results of the proposed model and training strategy is discussed. We compared the proposed semi-supervised network, i.e. the multi-task CNN trained on both datasets, with a fully-supervised network, i.e. a single-task CNN trained on only the AMC dataset. We conducted several experiments for comparing different configurations of the fully-supervised and semi-supervised networks.

\subsection{Experimental details}
The fully supervised network is composed of the encoder and the PET decoder and was trained on the AMC dataset in a fully supervised fashion, while the semi-supervised network further contains the ASL decoder and was trained by both AMC and LUMC datasets in a semi-supervised manner. 

The performance of the proposed model was compared with different configurations to study the effect of incorporating T1 scans as the input, the residual attention gate and the disentanglement attention gate on the results. The different flavors of the fully-supervised and semi-supervised network are denoted by $(+/-)$ T1, $(+/-)$ RA, $(+/-)$ DA which represent the network configurations with or without the T1w scan as the input, residual attention gate, and disentanglement attention gate, respectively. Note that in the fully-supervised CNN deploying the residual attention gate is not applicable, since the reconstructed ASL is not available there. 

In this study we compare the networks by using different metrics including SSIM, mean squared error (MSE) and Peak signal-to-noise ratio (PSNR). Since SSIM measures
image similarity using human perception aspects \cite{wang2004image}, we use it as the main score for comparing the networks. Also, a repeated measure one-way ANOVA test was performed on the SSIM values using a significance level of $p = 0.05$ between the proposed network and the other CNNs.

The number of the trainable parameters are 1307k, 1309k, 1445k, 1446k, for the networks with -T1-RA-DA, \mbox{+T1-RA-DA}, \mbox{+T1+RA-DA}, and \mbox{+T1+RA+DA} configurations.

\subsection{Quantitative results}
Table \ref{tab:mean_cross_valid} tabulates the average of SSIM, MSE and PSNR for the 7 folds on the paired dataset, i.e. AMC dataset only. Note that the metrics can only be computed for the paired data that includes ground truth, i.e. PET scan. 
Also, the comparative statistical significance of the networks with respect to the proposed CNN is shown by stars. In Section \ref{sec:observer_study} we will see that the semi-supervised network performs better in reconstructing the local changes.
Results show the fully-supervised CNNs achieved higher SSIM scores compare to the semi-supervised CNNs in general. However, the semi-supervised network with the \mbox{+T1+RA+DA} configuration yielded a comparable SSIM value to the fully-supervised networks. 

\begin{table*}[tb!]
\caption{The average results of 7-fold cross-validation for fully-supervised (single-task) and semi-supervised (multi-task) CNNs on the AMC dataset. S and M stand for single-task and multi-task networks. T1, RA, and DA denote whether or not the network has been trained by a T1w scan, a residual attention gate and a disentanglement attention gate, respectively. $\mu$ and $\sigma$ denote mean and standard deviation. The best SSIM values are shown in green.}
\begin{center}
\centering
\begin{tabular}{ccc|ccc|ccc|ccc}
\hline
\multirow{2}{*}{$\#$}
&\multirow{2}{*}{M/T}
&
&\multicolumn{3}{c|}{Hypercapnia}
&\multicolumn{3}{c|}{Normocapnia} 
&\multicolumn{3}{c}{All} \\
&&&SSIM& MSE &PSNR
&SSIM& MSE &PSNR  
&SSIM& MSE &PSNR  \\
\cline{4-12}
&
&
&$\mu\pm\sigma$&$\mu\pm\sigma$&$\mu\pm\sigma$
&$\mu\pm\sigma$&$\mu\pm\sigma$&$\mu\pm\sigma$ 
&$\mu\pm\sigma$&$\mu\pm\sigma$&$\mu\pm\sigma$\\
\hline
\Xhline{2\arrayrulewidth}
\Xhline{2\arrayrulewidth}
\multirow{8}{*}{Mean}
&\multirow{3}{*}{S}&{-T1-RA-DA} & 
0.81$\pm$0.09&0.02$\pm$0.01&19.6$\pm$4.6&
\textcolor{green}{0.87$\pm$0.08}&0.01$\pm$0.01&23.0$\pm$4.6&
0.85$\pm$0.09&0.01$\pm$0.01&21.7$\pm$4.9
\\
&&{+T1-RA-DA} & 
\textcolor{green}{0.82$\pm$0.09}&0.01$\pm$0.01&19.9$\pm$4.3&
\textcolor{green}{0.87$\pm$0.08}&0.01$\pm$0.01&23.3$\pm$4.3&
0.85$\pm$0.09&0.01$\pm$0.01&22.0$\pm$4.6
\\&&{+T1-RA+DA} & 
\textcolor{green}{0.82$\pm$0.09}&0.01$\pm$0.01&20.1$\pm$4.3&
\textcolor{green}{0.87$\pm$0.08}&0.01$\pm$0.01&23.4$\pm$4.3&
\textcolor{green}{0.85$\pm$0.08}&0.01$\pm$0.01&22.2$\pm$4.6
\\\cline{2-12}
&\multirow{4}{*}{M}
&
{-T1-RA-DA} & 
0.78$\pm$0.10 & 0.04$\pm$0.05 & 16.1$\pm$4.6&
0.83$\pm$0.09 &0.03$\pm$0.04  & 18.0$\pm$4.6&
0.81$\pm$0.10$^*$ & 0.03$\pm$0.04 & 17.3$\pm$4.7
\\
&
&{+T1-RA-DA} & 
0.78$\pm$0.10 & 0.05$\pm$0.09 & 17.4$\pm$6.4&
0.84$\pm$0.08 & 0.02$\pm$0.05 & 20.9$\pm$6.2&
0.82$\pm$0.09$^*$& 0.03$\pm$0.07 &19.6$\pm$6.5
\\
&&
{+T1+RA-DA} & 
{0.81$\pm$0.09} & 0.02$\pm$0.01 & 
19.7$\pm$4.4 &
{0.86$\pm$0.08} & 0.01$\pm$0.01&
23.0$\pm$4.5& {0.84$\pm$0.08}$^*$&
0.01$\pm$0.01 &
21.8$\pm$4.7
\\
&&{+T1+RA+DA} & 
\textcolor{green}{0.82$\pm$0.09} & 0.01$\pm$0.01 & 19.9$\pm$4.2&
\textcolor{green}{0.87$\pm$0.08} & 0.01$\pm$0.01 & 23.0$\pm$4.3&
\textcolor{green}{0.85$\pm$0.08}& 0.01$\pm$0.01 &21.8$\pm$4.5\\
\Xhline{2\arrayrulewidth}

\end{tabular}
\end{center}
\label{tab:mean_cross_valid}
\end{table*}

\subsection{Qualitative results}\label{sec:qualitative_res}

\begin{figure*}[tb!]
  \centering  
  \begin{tikzpicture}
    \node[anchor=south west,inner sep=0] at (0,6.)  {\includegraphics[width=18cm,trim={0 0cm 0 .8cm},clip]{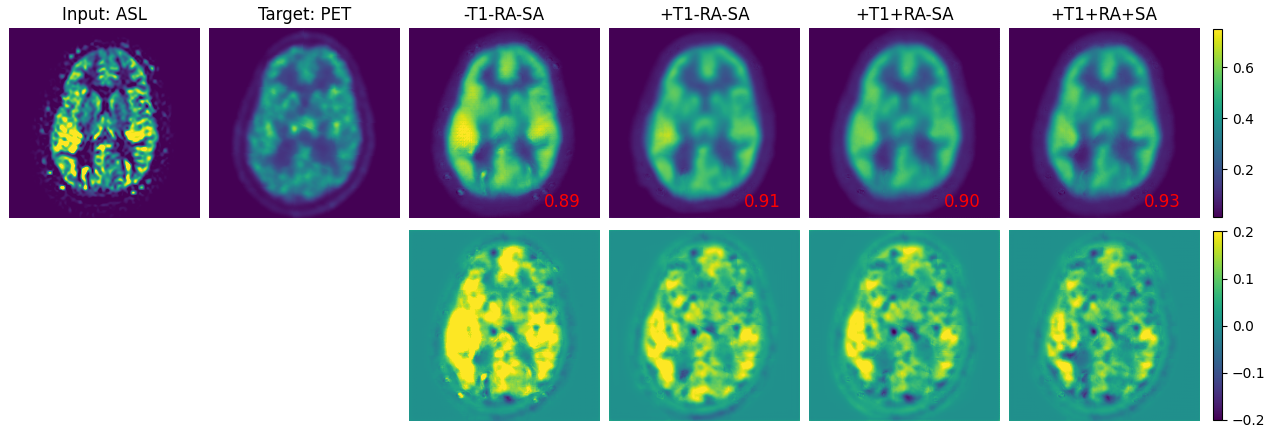}
    \put (-520,110) {\normalsize \rotatebox{90} {\small{Normocapnia}}}
    \put (-520,40) {\normalsize \rotatebox{90} {\small{Error}}}
    \put(-496,168){\small{Input: ASL}}
    \put(-420,168){\small{Target: PET}}
    \put(-335,168){\small{-T1-RA-DA}}
    \put(-260,168){\small{+T1-RA-DA}}
    \put(-180,168){\small{+T1+RA-DA}}
    \put(-100,168){\small{+T1+RA+DA}}
    };

    \node[anchor=south west,inner sep=0] at (0,0)  {\includegraphics[width=18cm,trim={.2cm 0cm .05cm .8cm}, clip]{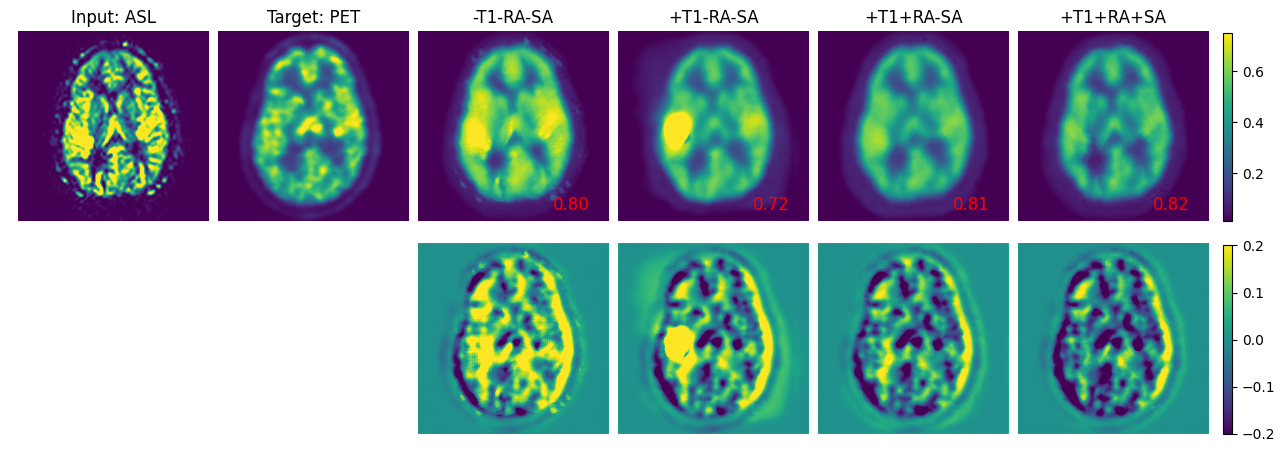}
    \put (-520,108) {\normalsize \rotatebox{90} {\small{Hypercapnia}}}
    \put (-520,35) {\normalsize \rotatebox{90} {\small{Error}}}};

\end{tikzpicture}
   \caption{Comparison between the output of the semi-supervised networks and the ground truth PET scan for a slice from the AMC dataset, for different network configurations. Top two rows show an example normocapnia result, and bottom two rows the same for the hypercapnia condition. -/+T1, -/+RA, -/+DA denote the networks with/without incorporating the T1w scan, a residual attention gate, and a disentanglement attention gate, respectively. The SSIM score for each output is reported in red. The hypercapnia scan shows a global change in blood flow in both ASL and PET scans.
   }   
  \label{fig:residual_effect} 
\end{figure*}
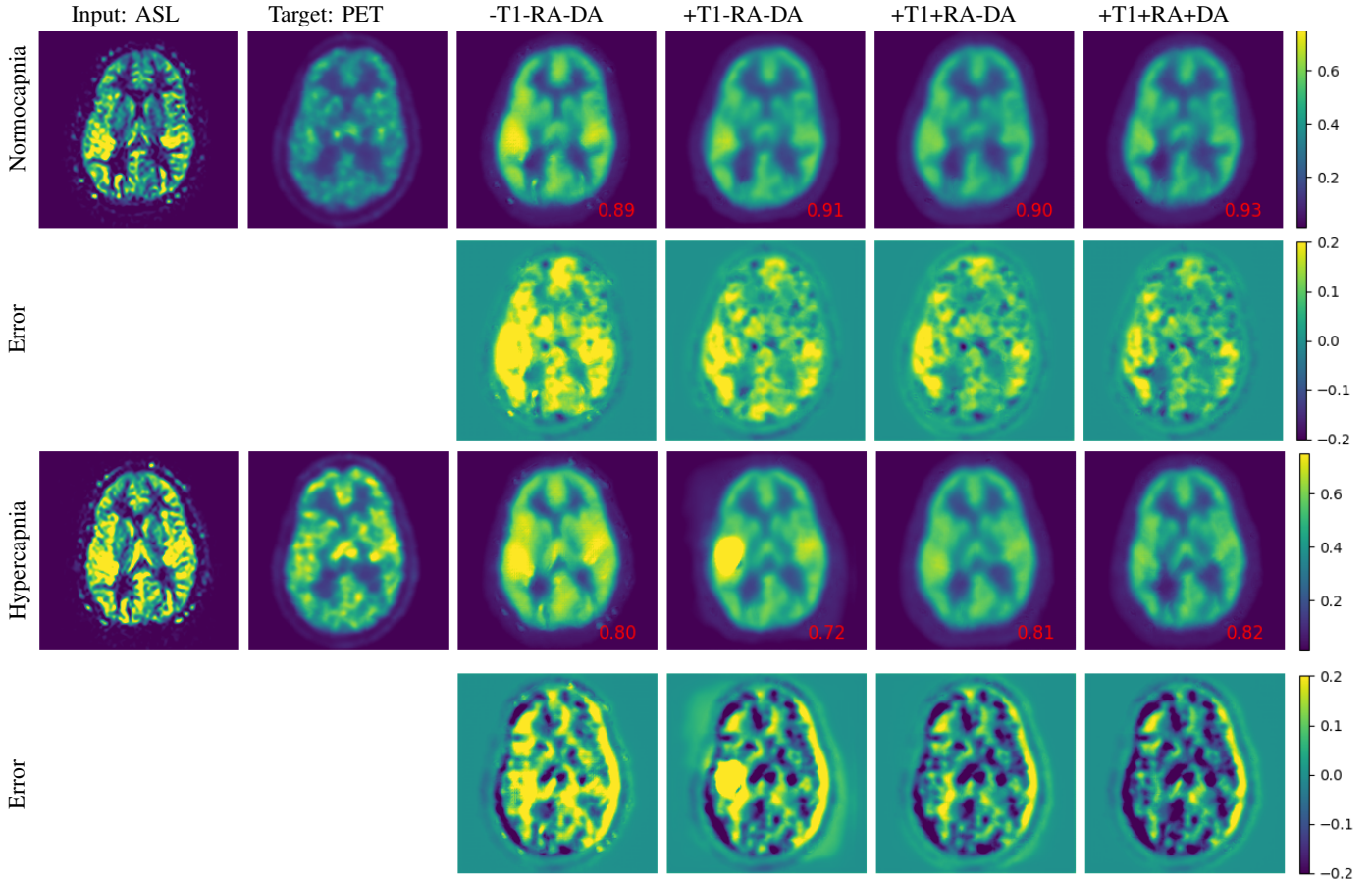

Figure \ref{fig:residual_effect} shows example results of different configurations of the proposed semi-supervised CNN, with/without incorporating the T1w scan, residual attention gate, and disentanglement attention gate, for one slice. As it can be seen, the ASL scan shows a higher blood flow in the insular cortex compare to the PET scan due to the dense presence of arteries in which slow flowing label is still present at the moment of readout. This caused a bias in synthesizing the details in the result of the -T1-RA-DA network. The result of +T1-RA-DA shows that including the T1w scan as a network input, alleviated this effect due to including anatomical structure in a high spatial resolution. Also, using the disentanglement attention and residual attention gates improved accuracy of the perfusion. 

\begin{figure*}[tb!]
  \centering  
  \begin{tikzpicture}
    \node[anchor=south west,inner sep=0] at (0,6.)  {\includegraphics[width=18cm,trim={0 0cm 0 .5cm},clip]{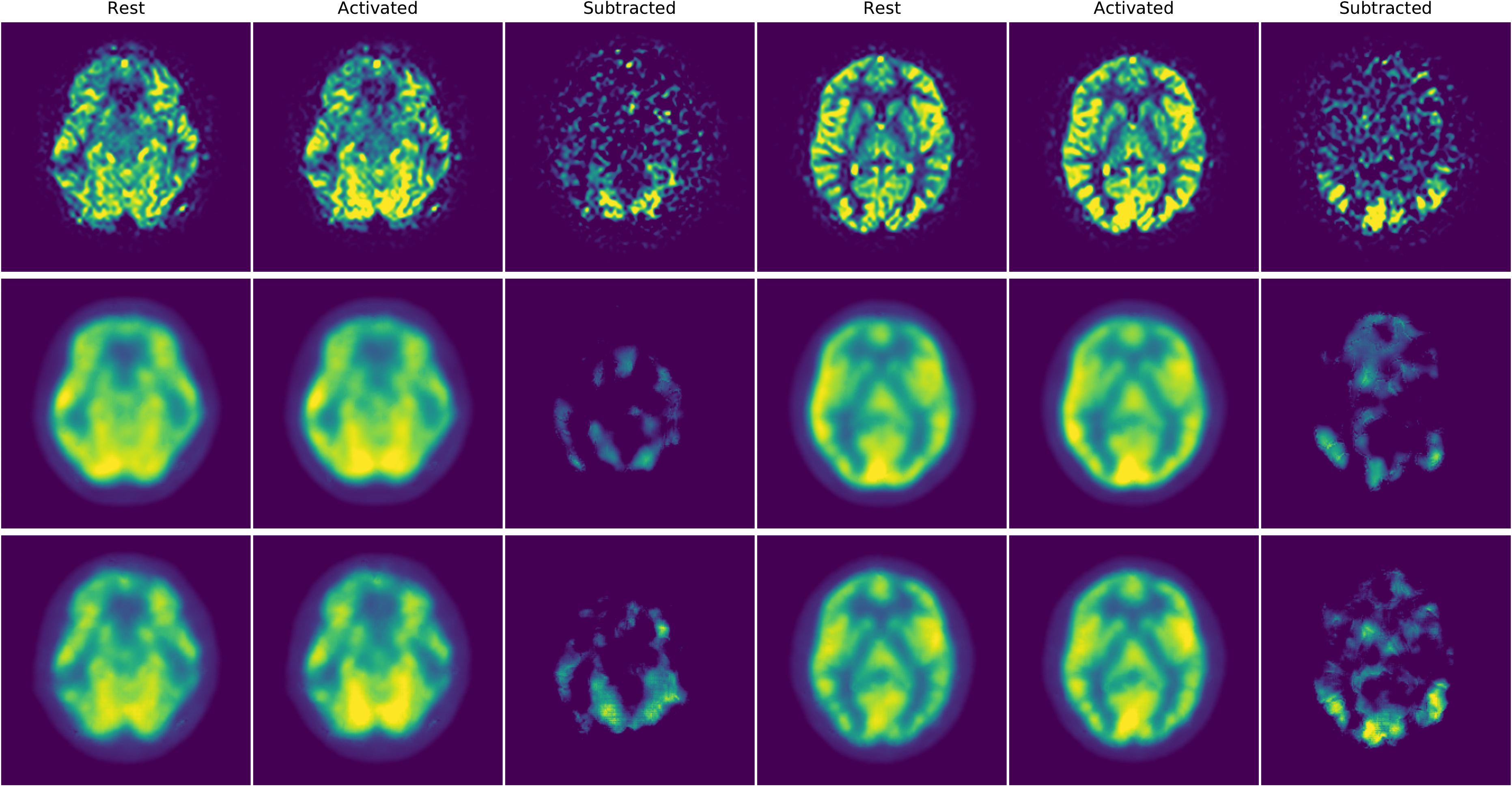}
    \put (-520,210) {\normalsize \rotatebox{90} {\small{ASL}}}
    \put (-520,100) {\normalsize \rotatebox{90} {\small{PET/ST CNN}}}
    \put (-520,20) {\normalsize \rotatebox{90} {\small{PET/MT CNN}}}
    \put(-480,263){\small{Rest}}
    \put(-410,263){\small{Activation}}
     \put(-405,279){\small{Slice $\#$1}}
    \put(-325,263){\small{Subtraction}}
    \put(-220,263){\small{Rest}}
    \put(-150,263){\small{Activation}}
    \put(-145,279){\small{Slice $\#$2}}
    \put(-70,263){\small{Subtraction}}
        };
   \draw [<->,pink,ultra thick] (0.1,15.6) -- (8.9,15.6);
   \draw [<->,pink,ultra thick] (9.1,15.6) -- (18.0,15.6);

    \end{tikzpicture}
   \caption{Comparison between the single-task (fully-supervised) and multi-task (semi-supervised) networks for two test samples from the visual cortex activation data (i.e. LUMC dataset). The slice $\#1$ is from the Tom $\&$ Jerry data, and the slice $\#2$ is from the checkerboard data. ST and MT stand for the single- and multi-task CNNs, respectively.
   }   
  \label{fig:visual_cortex} 
\end{figure*}

\begin{figure*}[tb!]
  \centering  
  \begin{tikzpicture}
    \node[anchor=south west,inner sep=0] at (0,6.)  {\includegraphics[width=18cm,trim={0 0cm 0 .5cm},clip]{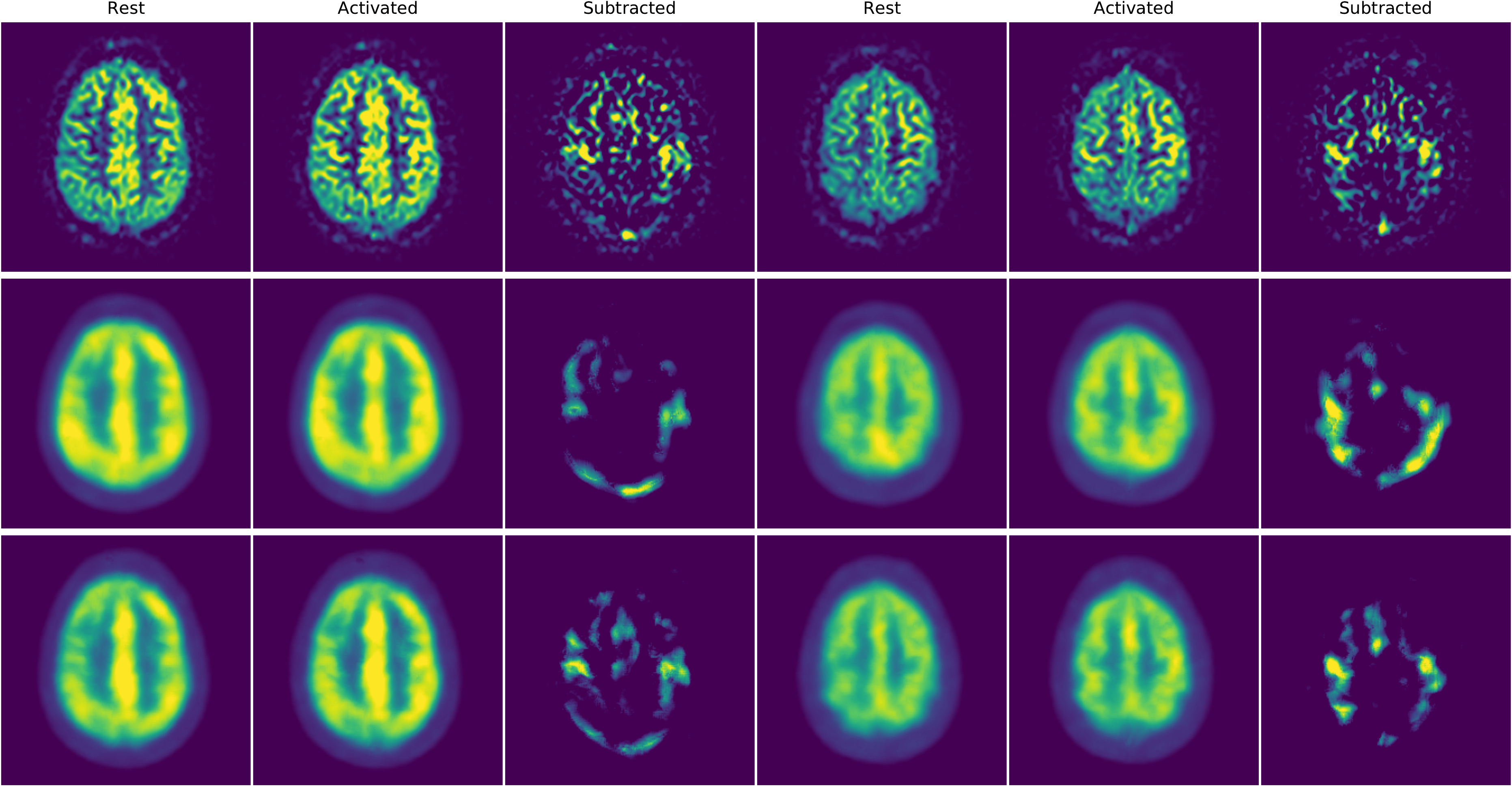}
    \put (-520,210) {\normalsize \rotatebox{90} {\small{ASL}}}
    \put (-520,100) {\normalsize \rotatebox{90} {\small{PET/ST CNN}}}
    \put (-520,20) {\normalsize \rotatebox{90} {\small{PET/MT CNN}}}
    \put(-480,263){\small{Rest}}
    \put(-410,263){\small{Activation}}
    \put(-325,263){\small{Subtraction}}
    \put(-405,279){\small{Slice $\#$1}}
    \put(-220,263){\small{Rest}}
    \put(-145,279){\small{Slice $\#$2}}
    \put(-150,263){\small{Activation}}
    \put(-70,263){\small{Subtraction}}
    };
     \draw [<->,pink,ultra thick] (0.1,15.6) -- (8.9,15.6);
   \draw [<->,pink,ultra thick] (9.1,15.6) -- (18.0,15.6);
    \end{tikzpicture}
   \caption{Comparison between the single-task (fully-supervised) and multi-task (semi-supervised) CNNs for two test slices from the motor cortex activation data (i.e. LUMC dataset). ST and MT stand for the single- and multi-task CNNs, respectively.
   }   
  \label{fig:motor_cortex} 
\end{figure*}

Figure \ref{fig:visual_cortex} and \ref{fig:motor_cortex} show the fully-supervised and semi-supervised CNN outputs for the visual cortex and motor cortex activated data (i.e. LUMC data), respectively. The subtracted scan is obtained by subtracting the rest scan from the activation scan.  According to these images the semi-supervised network performs better in locating and synthesizing the activated regions. 

\subsection{Observer study}\label{sec:observer_study}
Due to the absence of PET scans in the unpaired dataset, i.e. the LUMC dataset, a quantitative evaluation is not possible. Therefore, we conducted a randomized double-blind observer study for judging the results of the fully-supervised and semi-supervised networks on this dataset. For this goal, a comparison was performed for the reconstructed PET scans of the fully-supervised CNN with the +T1-RA+DA configuration and the semi-supervised CNN with the +T1+RA+DA configuration. The quality of the results were scored by two observers on a 5-point scale (1 is worst and 5 is best) for the overall reconstructed image as well as how well the activated region was identified. The activated regions were judged by comparing the activated CBF map of the reconstructed PET images to the subtraction of ASL activated and rest map. The observers were provided with the fully-supervised and semi-supervised reconstructed PET scans in both rest and activation conditions, the input ASL scan in both rest and activation conditions and the subtraction of the rest scans from the corresponding activation scan for both reconstructed PET and input ASL scans. The measuring process was performed by considering the physiological information and textural structure, including contrast, artifacts, sharpness, and structure of the reconstructed and subtracted PET scans compared to the input and subtracted ASL scans. The results are tabulated in Table \ref{tbl:observer_study} for the observers and three different activation maps including the visual cortex, and the motor cortex. According to the average scores, the semi-supervised network performed better for reconstructing the motor-cortex activation data, i.e. finger tapping, in terms of both the entire PET scan and the activated area. Also, the semi-supervised network for reconstructing the visual-cortex activation works better for the Tom $\&$ Jerry data in terms of both the entire PET scan and activated area and for the checkerboard data in terms of the entire PET scan.

\begin{table*}[tbh!]
\caption{Results of the observer study for the test dataset from the LUMC dataset. Activation stands for the activated region only which is obtained by subtracting the rest scan from the activated scan. $\mu$ and $\sigma$ denote mean and standard deviation.}
\begin{center}
\begin{tabular}{ |cc|cc|cc|cc||cc| } 
 \hline
 \multirow{3}{*}{Observers} && \multicolumn{2}{c|}{Tom $\&$ Jerry} & \multicolumn{2}{c|}{Checkerboard} & \multicolumn{2}{c||}{Motor cortex}&\multicolumn{2}{c|}{Average}\\
 && PET & Activation  & PET & Activation  & PET & Activation &PET & Activation  \\\cline{3-10}
   && $\mu \pm \sigma$ &$\mu \pm \sigma$ & $\mu \pm \sigma$  & $\mu \pm \sigma$ &$\mu \pm \sigma$ &$\mu \pm \sigma$ &$\mu \pm \sigma$&$\mu \pm \sigma$ \\
 \hline \Xhline{2\arrayrulewidth}
 \multirow{2}{*}{$\#1$} & S & 3.3 $\pm$ 0.5 & 2.7 $\pm$ 0.9&  3.1 $\pm$ 0.8& \textbf{2.7} $\pm$ \textbf{0.9}&  3.0 $\pm$ 0.5& 2.3 $\pm$ 0.7&3.1 $\pm$ 0.6& 2.6 $\pm$ 0.9\\ 
                        & M & \textbf{3.4} $\pm$ \textbf{0.5}& \textbf{3.2} $\pm$ \textbf{0.9} & \textbf{3.3} $\pm$ \textbf{0.6}& 2.6 $\pm$ 1.2& \textbf{3.3} $\pm$ \textbf{0.5}& \textbf{3.4} $\pm$ 0.8& \textbf{3.3} $\pm$ \textbf{0.5}& \textbf{3.1} $\pm$ \textbf{1.0}\\ \hline
\multirow{2}{*}{$\#2$} &S&1.8 $\pm$ 0.7&1.1 $\pm$ 0.4&1.8 $\pm$ 0.7&1.2 $\pm$ 0.4& 1.8 $\pm$ 0.6&1.3$\pm$ 0.4&1.8 $\pm$ 0.7&1.2 $\pm$ 0.4\\
 &M&\textbf{1.9} $\pm$ \textbf{0.6}&\textbf{1.4} $\pm$ \textbf{0.5}&\textbf{1.9} $\pm$ \textbf{0.5}&1.2 $\pm$ 0.4&\textbf{1.9} $\pm$ \textbf{0.6} &\textbf{1.4} $\pm$ \textbf{0.5}&\textbf{1.9} $\pm$ \textbf{0.6}&\textbf{1.3} $\pm$ \textbf{0.5}\\
 \hline\hline
 \multirow{2}{*}{Average} &S&2.6 $\pm$ 1.0&1.8 $\pm$ 1.0&2.5 $\pm$ 1.0& 1.7 $\pm$ 1.0& 2.4 $\pm$ 0.8&1.7 $\pm$ 0.8& 2.4 $\pm$ 0.9 & 1.7 $\pm$ 0.9\\
 &M&\textbf{2.7} $\pm$ \textbf{0.9}&\textbf{2.2} $\pm$ \textbf{1.1}&\textbf{2.6} $\pm$ \textbf{0.9}&1.7 $\pm$ 1.0 &\textbf{2.6} $\pm$ \textbf{0.9}&\textbf{2.2} $\pm$ \textbf{1.2}&\textbf{2.6} $\pm$ \textbf{0.9}&\textbf{2.0} $\pm$ \textbf{1.1}\\
 \hline
\end{tabular}\label{tbl:observer_study}
\end{center}
\end{table*}
\section{Discussion}
In this paper we proposed a new semi-supervised CNN in a multi-task fashion for translating ASL and T1-weighted scans to PET scans. Since it is difficult to acquire substantial amounts of PET scans, we leveraged a semi-supervised learning mechanism to make the network capable of learning physiological features from paired with PET scans and unpaired data.
For this goal, we designed a CNN composed of one encoder and two decoders. The encoder extracts features from ASL and T1 scans, while the PET decoder synthesizes a PET scan and the ASL decoder reconstructs ASL scan. The PET and the ASL scans are generated  with and without the aid of the skip connections, respectively. Deploying the skip connections for the PET decoder assists the network in recovering the image resolution, while nonexistence of the skip connections for the ASL encourages the encoder to extract the relevant physiological features as much as possible. Also, we utilized channel attention gates for each skip connection between the encoder and PET decoder in order to disentangle the ASL and PET features. 

For each batch, an alternating training scheme including coarse and fine-grained learning was performed. In the first step, ASL and T1w scans are fed to the network to synthesize PET and reconstruct ASL. 
In the second step, a residual attention map is calculated by a spatial attention gate and re-fed to the network to encourage it to concentrate on the fine details. 

We compared the proposed method with a traditional fully-supervised network (and thus only trained on the paired PET-MRI data) in a single-task fashion. A 7-fold cross-validation was performed for comparing the CNNs with four different configurations: \mbox{-T1-RA-DA}, \mbox{+T1-RA-DA}, \mbox{+T1+RA-DA}, \mbox{+T1+RA+DA} and three fully-supervised configurations composed of: \mbox{-T1-RA-DA}, \mbox{+T1-RA-DA} and \mbox{+T1-RA+DA}, in which $(+/-)$ T1, $(+/-)$ RA, $(+/-)$ DA which represent the network configurations with or without the T1w scan as the input, residual attention gate, and disentanglement attention gate, respectively.

According to Table \ref{tab:cross_valid}, in 5 out of 7 folds both the fully-supervised and semi-supervised networks performed equal or better when including the T1w scan as an input (i.e.\mbox{+T1-RA-DA}), in terms of SSIM score. Apparently, the incorporation of high spatial resolution information aided the networks to extract the geometrical structure, which was beneficial for predicting the PET. This is not unexpected, since the partial volume percentage of gray matter in a voxel of a perfusion scan is the main driver of the signal strength in perfusion images \cite{asllani2008regression}. This does entail, however, also a risk: when the network would rely too much on anatomical information as opposed to physiological input, important information can be lost. This was one of the main reasons why we also included the validation step in which neuronal activation was used to locally change the physiological state. As seen in Figure \ref{fig:visual_cortex} and Figure \ref{fig:motor_cortex}, the network was still able to identify the activated regions correctly, showing that the network did not just rely on the structural information. 

\begin{table*}[thb!]
\caption{Results of 7-fold cross-validation for fully-supervised, (single-task), and semi-supervised (multi-task), CNNs on AMC dataset. The average results of all folds are reported in Table \ref{tab:mean_cross_valid}.  T1, RA, DA denote whether or not the network has been trained by T1-weighted scan, residual attention gate and disentanglement attention gate, respectively. The best SSIM value in each fold is shown in green. The comparative statistical significance of the networks with respect to the proposed CNN is shown by stars.}
\begin{center}

\centering
\begin{tabular}{ccc|ccc|ccc|ccc}
\hline
\multirow{2}{*}{$\#$}
&\multirow{2}{*}{M/T}
&
&\multicolumn{3}{c|}{Hypercapnia}
&\multicolumn{3}{c|}{Normocapnia} 
&\multicolumn{3}{c}{All} \\
&&&SSIM& MSE &PSNR
&SSIM& MSE &PSNR  
&SSIM& MSE &PSNR  \\

\hline
\multirow{7}{*}{1}
&
\multirow{3}{*}{S}&{-T1-RA-DA} & 
0.80$\pm$0.13&0.02$\pm$0.02&20.5$\pm$5.2 &
0.86$\pm$0.11&0.01$\pm$0.01&23.2$\pm$4.7&
0.84$\pm$0.12&0.01$\pm$0.01&22.1$\pm$5.1
\\&&{+T1-RA-DA} & 
\textcolor{green}{0.81$\pm$0.13}&0.01$\pm$0.02&20.9$\pm$5.1&
\textcolor{green}{0.86$\pm$0.11}&0.01$\pm$0.01&23.4$\pm$4.4&
\textcolor{green}{0.84$\pm$0.12}$^*$&0.01$\pm$0.01&22.4$\pm$4.9
\\
&&{+T1-RA+DA} & 
0.80$\pm$0.11&0.01$\pm$0.01&21.0$\pm$4.8&
0.84$\pm$0.10&0.01$\pm$0.01&23.4$\pm$4.3&
0.82$\pm$0.10&0.01$\pm$0.01&22.4$\pm$4.7
\\
\cline{2-12}
&\multirow{4}{*}{M}
&
{-T1-RA-DA} & 
0.73$\pm$0.14 & 0.08$\pm$0.09 & 15.7$\pm$7.3&
0.80$\pm$0.12 & 0.04$\pm$0.07 & 19.4$\pm$7.6&
0.78$\pm$0.13$^*$ & 0.06$\pm$0.08 & 17.9$\pm$7.7
\\&&
{+T1-RA-DA} & 
0.77$\pm$0.11 & 0.01$\pm$0.01 & 20.5$\pm$4.7&
0.83$\pm$0.10 & 0.01$\pm$0.01 & 22.7$\pm$4.4&
{0.80$\pm$0.11}$^*$ & 0.01$\pm$0.01 & 21.8$\pm$4.7
\\&&
{+T1+RA-DA} & 
{0.80$\pm$0.12} & 0.01$\pm$0.01 & 20.6$\pm$5.0&
{0.85$\pm$0.11} & 0.01$\pm$0.01 & 23.3$\pm$4.6&
{0.83$\pm$0.12} & 0.01$\pm$0.01 & 22.3$\pm$4.9
\\
&&{+T1+RA+DA} & 
{0.80$\pm$0.13} & 0.02$\pm$0.02 & 20.4$\pm$4.7&
{0.85$\pm$0.11} & 0.01$\pm$0.01 & 22.4$\pm$4.5&
{0.83$\pm$0.12} & 0.01$\pm$0.01 & 21.6$\pm$4.7
\\

\Xhline{2\arrayrulewidth}
\multirow{7}{*}{2}
&\multirow{3}{*}{S}&{-T1-RA-DA} & 
0.82$\pm$0.10&0.010$\pm$0.01&20.4$\pm$4.4&
0.86$\pm$0.09&0.01$\pm$0.01&23.7$\pm$4.2&
0.84$\pm$0.09&0.01$\pm$0.01&22.3$\pm$4.6
\\&&{+T1-RA-DA} & 
0.81$\pm$0.10&0.01$\pm$0.01&20.4$\pm$4.5&
0.86$\pm$0.09&0.01$\pm$0.01&23.6$\pm$4.2&
0.84$\pm$0.10$^*$&0.01$\pm$0.01&22.3$\pm$4.6
\\&&{+T1-RA+DA} & 
0.81$\pm$0.10&0.01$\pm$0.01&20.3$\pm$4.3&
0.86$\pm$0.09&0.01$\pm$0.01&23.3$\pm$4.1&
0.84$\pm$0.10&0.01$\pm$0.01&22.0$\pm$4.4
\\
\cline{2-12}
&\multirow{4}{*}{M}
&
{-T1-RA-DA} & 
0.80$\pm$0.10 & 0.02$\pm$0.03 & 18.8$\pm$4.9&
0.83$\pm$0.10 & 0.03$\pm$0.04 & 20.7$\pm$6.4&
0.81$\pm$0.11$^*$ & 0.02$\pm$0.04 & 19.9$\pm$5.9

\\
&&
{+T1-RA-DA} & 
{0.81$\pm$0.10} & 0.02$\pm$0.01 & 19.8$\pm$4.3&
0.85$\pm$0.10 & 0.01$\pm$0.01 & 22.9$\pm$4.3&
0.83$\pm$0.10$^*$ & 0.01$\pm$0.01& 21.6$\pm$4.6
\\
&&
{+T1+RA-DA} & 
{0.81$\pm$0.10}& 0.01$\pm$0.01 & 20.4$\pm$4.3&
{0.86$\pm$0.09} & 0.01$\pm$0.01 & 23.6$\pm$4.2&
{0.84$\pm$0.09}$^*$ & 0.01$\pm$0.01 & 22.2$\pm$4.5
\\
&&{+T1+RA+DA} & 
\textcolor{green}{0.82$\pm$0.09}&0.01$\pm$0.01&24.4$\pm$4.5&
\textcolor{green}{0.87$\pm$0.09}&0.01$\pm$0.01&23.7$\pm$4.1&
\textcolor{green}{0.85$\pm$0.09} & 0.01$\pm$0.01 & 22.3$\pm$4.6\\

\Xhline{2\arrayrulewidth}
\multirow{8}{*}{3}
&\multirow{3}{*}{S}&{-T1-RA-DA} & 
0.83$\pm$0.08&0.02$\pm$0.01&20.0$\pm$4.5&
0.87$\pm$0.07&0.01$\pm$0.01&23.1$\pm$4.9&
0.86$\pm$0.08&0.01$\pm$0.01&22.0$\pm$5.1
\\&&{+T1-RA-DA} & 
0.84$\pm$0.07&0.01$\pm$0.01&20.2$\pm$4.6&
0.88$\pm$0.06&0.01$\pm$0.01&23.3$\pm$4.4&
0.86$\pm$0.07$^*$&0.01$\pm$0.01&22.3$\pm$4.7
\\&&{+T1-RA+DA} & 
0.84$\pm$0.07&0.02$\pm$0.01&20.0$\pm$4.9&
0.88$\pm$0.06&0.01$\pm$0.01&23.1$\pm$4.6&
0.86$\pm$0.07&0.01$\pm$0.01&22.0$\pm$4.9
\\
\cline{2-12}
&
\multirow{4}{*}{M}&
{-T1-RA-DA} & 
0.81$\pm$0.08 & 0.03$\pm$0.01 & 15.8$\pm$1.8&
0.85$\pm$0.07 & 0.02$\pm$0.01 & 16.7$\pm$1.2&
{0.84$\pm$0.07}$^*$ & 0.02$\pm$0.01 & 16.4$\pm$1.5

\\
&&
{+T1-RA-DA} & 
{0.81$\pm$0.07} & 0.01$\pm$0.01 & 20.2$\pm$4.4&
0.84$\pm$0.05 & 0.01$\pm$0.00 & 23.0$\pm$4.0&
0.83$\pm$0.06$^*$ & 0.01$\pm$0.01& 22.0$\pm$4.3 
\\
&&
{+T1+RA-DA} & 
{0.81$\pm$0.07} & 0.02$\pm$0.02 & 18.2$\pm$4.8&
{0.85$\pm$0.06} & 0.01$\pm$0.01 & 21.3$\pm$4.8&
{0.84$\pm$0.07}$^*$ & 0.02$\pm$0.01 & 20.3$\pm$5.0
\\
&&{+T1+RA+DA} & 
\textcolor{green}{0.85$\pm$0.07} & 0.01$\pm$0.01 & 20.6$\pm$4.6&
\textcolor{green}{0.88$\pm$0.06} & 0.01$\pm$0.00 & 23.4$\pm$4.1&
\textcolor{green}{0.87$\pm$0.06} & 0.01$\pm$0.01 & 22.5$\pm$4.5

\\
\Xhline{2\arrayrulewidth}
\multirow{8}{*}{4}
&\multirow{3}{*}{S}&{-T1-RA-DA} & 
0.80$\pm$0.07&0.02$\pm$0.02&18.7$\pm$4.5&
0.90$\pm$0.05&0.01$\pm$0.01&24.9$\pm$4.5&
0.87$\pm$ 0.07&0.01$\pm$0.01&23.3$\pm$5.2
\\&&{+T1-RA-DA} & 
\textcolor{green}{0.80$\pm$0.07}&0.02$\pm$0.02&18.5$\pm$4.5&
\textcolor{green}{0.90$\pm$0.05}&0.00$\pm$0.00&25.5$\pm$3.8&
\textcolor{green}{0.87$\pm$0.07}&0.01$\pm$0.01&23.8$\pm$5.1
\\&&{+T1-RA+DA} & 
0.80$\pm$0.08&
0.02$\pm$0.02&
19.1$\pm$4.9&
0.90$\pm$0.06&
0.00$\pm$0.00&
25.7$\pm$3.8&
0.87$\pm$0.08&
0.01$\pm$0.01&
24.0$\pm$5.0\\
\cline{2-12}
&
\multirow{4}{*}{M}&
{-T1-RA-DA} & 
0.78$\pm$0.06 & 0.03$\pm$0.02 & 15.1$\pm$1.9&
{0.88$\pm$0.05} & 0.02$\pm$0.00 & 17.4$\pm$0.9&
{0.86$\pm$0.07} & 0.02$\pm$0.01 & 16.8$\pm$1.6
\\
&&
{+T1-RA-DA} & 
{0.79$\pm$0.07} & 0.03$\pm$0.03 & 18.1$\pm$5.1&
0.88$\pm$0.06 & 0.02$\pm$0.05 & 22.7$\pm$6.5&
{0.86$\pm$0.07} & 0.02$\pm$0.05& 21.6$\pm$6.5 
\\
&&
{+T1+RA-DA} & 
\textcolor{green}{0.80$\pm$0.07} & 0.02$\pm$0.02 & 18.8$\pm$5.0&
\textcolor{green}{0.90$\pm$0.05} & 0.00$\pm$0.00 & 25.6$\pm$3.8&
\textcolor{green}{0.87$\pm$0.07} & 0.01$\pm$0.01 & 23.9$\pm5.1$
\\ 
&&{+T1+RA+DA} & 
\textcolor{green}{0.80$\pm$0.07} & 0.02$\pm$0.02 & 18.6$\pm$4.5&
\textcolor{green}{0.90$\pm$0.05}&0.00$\pm$0.00& 25.3$\pm$3.7&
\textcolor{green}{0.87$\pm$0.07}&0.01$\pm$0.01&23.6$\pm$4.9\\

\Xhline{2\arrayrulewidth}
\multirow{8}{*}{5}
&\multirow{3}{*}{S}&{-T1-RA-DA} & 
0.80$\pm$0.1&0.01$\pm$0.01&20.0$\pm$4.0&
0.85$\pm$0.09&0.01$\pm$0.01&228$\pm$3.9&
0.83$\pm$0.09&0.01$\pm$0.01&21.6$\pm$4.2
\\&&{+T1-RA-DA} & 
0.80$\pm$0.10&0.01$\pm$0.01&20.0$\pm$3.9&
0.85$\pm$0.08&0.01$\pm$0.01&22.8$\pm$3.6&
0.83$\pm$0.09&0.01$\pm$0.09&21.7$\pm$4.1
\\&&{+T1-RA+DA} & 
0.81$\pm$0.09&0.01$\pm$0.01&20.4$\pm$3.9&
0.86$\pm$0.08&0.01$\pm$0.01&23.3$\pm$3.8&
0.84$\pm$0.09&0.01$\pm$0.01&22.1$\pm$4.1
\\
\cline{2-12}
&
\multirow{4}{*}{M}&
{-T1-RA-DA} & 
0.75$\pm$0.11 & 0.05$\pm$0.07 & 16.9$\pm$6.0&
0.80$\pm$0.09 & 0.03$\pm$0.05 & 19.7$\pm$5.7&
0.78$\pm$0.10$^*$ & 0.03$\pm$0.05 & 18.6$\pm$6.0

\\
&
&{+T1-RA-DA} & 
0.75$\pm$0.11 & 0.08$\pm$0.09 & 14.6$\pm$6.9&
0.80$\pm$0.09 & 0.06$\pm$0.08 & 17.4$\pm$7.1&
0.78$\pm$0.11$^*$ & 0.07$\pm$0.09& 16.3$\pm$7.2
\\
&&
{+T1+RA-DA} & 
{0.79$\pm$0.08} & 0.01$\pm$0.01 & 20.0$\pm$3.7&
{0.84$\pm$0.07} & 0.01$\pm$0.01 & 22.6$\pm$3.5&
{0.82$\pm$0.08}$^*$ & 0.01$\pm$0.01 & 21.6$\pm$3.9
\\
&&{+T1+RA+DA} & 
\textcolor{green}{0.80$\pm$0.09} & 0.01$\pm$0.01 & 20.2$\pm$3.9&
\textcolor{green}{0.85$\pm$0.08} & 0.01$\pm$0.01 & 22.9$\pm$3.7&
\textcolor{green}{0.83$\pm$0.09} & 0.01$\pm$0.01 & 21.8$\pm$4.0

\\
\Xhline{2\arrayrulewidth}
\multirow{8}{*}{6}
&\multirow{3}{*}{S}&{-T1-RA-DA} & 
0.82$\pm$0.07&0.02$\pm$0.01&18.3$\pm$4.8&
0.87$\pm$0.06&0.01$\pm$0.01&21.5$\pm$5.0&
0.85$\pm$0.07&0.02$\pm$0.01&20.2$\pm$5.2
\\&&{+T1-RA-DA} & 
\textcolor{green}{0.84$\pm$0.06}&0.02$\pm$0.01&19.4$\pm$4.1&
0.87$\pm$0.06&0.01$\pm$0.01&22.0$\pm$5.0&
\textcolor{green}{0.86$\pm$0.06}&0.01$\pm$0.01&21.0$\pm$4.8
\\&&{+T1-RA+DA} & 
0.84$\pm$0.05&0.01$\pm$0.01&19.7$\pm$4.2&
0.87$\pm$0.05&0.01$\pm$0.01&22.2$\pm$4.8&
0.86$\pm$0.05&0.01$\pm$0.01&21.2$\pm$4.7
\\
\cline{2-12}
&\multirow{4}{*}{M}
&
{-T1-RA-DA} & 
0.78$\pm$0.07 & 0.03$\pm$0.01 & 14.9$\pm$1.7&
0.83$\pm$0.06 & 0.03$\pm$0.01 & 16.2$\pm$1.6&
0.81$\pm$0.07$^*$ & 0.03$\pm$0.01& 15.7$\pm$1.7 
\\
&&
{+T1-RA-DA} & 
0.77$\pm$0.11 & 0.14$\pm$0.15 & 13.2$\pm$7.8&
0.85$\pm$0.07 & 0.04$\pm$0.05 & 18.6$\pm$6.6&
0.82$\pm$0.10$^*$ & 0.08$\pm$0.11& 16.4$\pm$7.6
\\
&&
{+T1+RA-DA} & 
{0.82$\pm$0.06} & 0.02$\pm$0.01 & 19.3$\pm$4.2&
{0.85$\pm$0.05} & 0.01$\pm$0.01 & 21.9$\pm$4.7&
{0.84$\pm$0.06}$^*$ & 0.01$\pm$0.01 & 20.9$\pm$4.7
\\
&&{+T1+RA+DA} & 
0.83$\pm$0.06& 0.02$\pm$0.01 & 19.4$\pm$4.3&
\textcolor{green}{0.88$\pm$0.06} & 0.01$\pm$0.01 & 21.8$\pm$4.9&
\textcolor{green}{0.86$\pm$0.06} & 0.01$\pm$0.01 & 20.8$\pm$4.8

\\
\Xhline{2\arrayrulewidth}
\multirow{8}{*}{7}
&\multirow{3}{*}{S}&{-T1-RA-DA} & 
0.83$\pm$0.06&0.02$\pm$0.01&19.4$\pm$3.8&
0.86$\pm$0.06&0.01$\pm$0.01&22.1$\pm$4.3&
0.85$\pm$0.06&0.01$\pm$0.01&21.0$\pm$4.3
\\
&&{+T1-RA-DA} & 
\textcolor{green}{0.82$\pm$0.06}&0.01$\pm$0.01&19.6$\pm$3.4&
\textcolor{green}{0.86$\pm$0.07}&0.01$\pm$0.00&22.5$\pm$3.9&
\textcolor{green}{0.85$\pm$0.07}$^*$&0.01$\pm$0.01&21.4$\pm$3.9
\\&&{+T1-RA+DA} & 
0.82$\pm$0.06&0.01$\pm$0.01&19.7$\pm$3.4&
0.85$\pm$0.07&0.01$\pm$0.01&23.0$\pm$3.9&
0.84$\pm$0.07&0.01$\pm$0.01&21.6$\pm$4.0
\\\cline{2-12}
&\multirow{4}{*}{M}
&
{-T1-RA-DA} & 
{0.80$\pm$0.06} & 0.03$\pm$0.01 & 15.6$\pm$1.4&
0.83$\pm$0.06 & 0.02$\pm$0.01 & 16.6$\pm$1.1&
0.82$\pm$0.07$^*$ & 0.03$\pm$0.01 & 16.2$\pm$1.3
\\
&&
{+T1-RA-DA} & 
0.80$\pm$0.08 & 0.04$\pm$0.05 & 17.2$\pm$5.5&
{0.84$\pm$0.07} & 0.03$\pm$0.04 & 19.7$\pm$6.0&
{0.83$\pm$0.08}$^*$ & 0.03$\pm$0.05 & 18.7$\pm$5.9
\\
&&
{+T1+RA-DA} & 
\textcolor{green}{0.82$\pm$0.06} & 0.01$\pm$0.01 & 19.7$\pm$3.4&
{0.85$\pm$0.07} & 0.01$\pm$0.00 & 22.7$\pm$3.8&
{0.84$\pm$0.07} & 0.01$\pm$0.01 & 21.5$\pm$3.9
\\
&&{+T1+RA+DA} & 
\textcolor{green}{0.82$\pm$0.06} & 0.01$\pm$0.01 & 19.5$\pm$3.3&
{0.85$\pm$0.07} & 0.01$\pm$0.01 & 22.0$\pm$4.0&
{0.84$\pm$0.07} & 0.01$\pm$0.01 & 21.0$\pm$3.9
\\

\Xhline{2\arrayrulewidth}
\end{tabular}
\end{center}
\label{tab:cross_valid}
\end{table*}

In 4 out of 7 folds, the fully-supervised CNN with the disentanglement attention gates (i.e.\mbox{+T1-RA+DA}) performed better than or equal to the network without deploying attention gate (i.e.\mbox{+T1-RA-DA}). This is probably due to the ability of such a gate to selectively filter out irrelevant features. This configuration also performed the best among the fully-supervised networks when looking at the average SSIM score, see Table \ref{tab:mean_cross_valid}.

In all folds (7 out of 7), the incorporation of the residual attention gate in combination with the alternating training scheme improved the results, {cf.} \mbox{+T1+RA-DA} with \mbox{+T1-RA-DA} for the semi-supervised CNNs in Table \ref{tab:cross_valid}. This improvement is probably due to the ability of the residual attention gate to focus on discrepant fine details between the reconstructed PET scan and the ground truth which cause a high loss value. Also in all folds, incorporating the residual and disentanglement attention gates improved the results, {cf.} \mbox{+T1+RA+DA} with \mbox{+T1+RA-DA} for the semi-supervised CNN. This is due to combination of the aforementioned benefits of incorporating the T1-weighted scan as the CNN's input, and deploying the disentanglement and residual attention gates.

Comparing the best semi-supervised network, i.e. \mbox{+T1+RA-DA}, with the fully-supervised networks showed that in 2 out of 7 and 5 out of 7 folds the \mbox{+T1+RA-DA} network worked significantly and not significantly better than or equal to the fully-supervised CNN with incorporating T1 scan (i.e. \mbox{+T1-RA-DA}). 

On average, the fully-supervised CNN performed better than the semi-supervised CNN without incorporating the residual attention gate in terms of the SSIM score (see Table \ref{tab:mean_cross_valid}) on reconstructing PET scans when we have a rest scan or global activation changes. As mentioned before, this is due to training the fully-supervised network by the paired data, which only contain the global activation changes, whereas the semi-supervised network was trained on a broader dataset by including both paired and unpaired data, which include both local and global activation changes. This leads the semi-supervised network to represent a wider spectrum of possible cases, while the fully-supervised CNN has a simpler task by concentrating on the global changes. However, deploying the residual attention gate incorporated with the disentanglement attention gates aided the semi-supervised CNN to alleviate this distraction. This can be seen by the results of the semi-supervised CNN with the \mbox{+T1+RA+DA} configuration which yielded comparable results to the fully-supervised CNNs in terms of the SSIM score.
Also, we conducted an observer study to evaluate the results of the best configurations of the fully-supervised vs. semi-supervised networks, i.e. \mbox{+T1-RA+DA} vs. \mbox{+T1+RA+DA}, on the LUMC dataset. The scoring was performed blindly by two observers. The scores showed semi-supervised CNN performing better in synthesizing the local activation changes, i.e. visual and motor cortex activation data (see Table \ref{tbl:observer_study}). We can conclude that the semi-supervised CNN with the \mbox{+T1+RA+DA} configuration performed the best on both local and global activation changes. This is due to  deploying the proposed alternating training strategy and the training the network on a wider dataset. 

There are some limitations in this study which can be addressed in future works. First, to further enhance the robustness of the proposed network, we consider to increase the training data, both with global and local changes. Second, in the present study the activated data is limited to visual and motor cortex and no patient data has been included in the dataset. Studying and evaluating the proposed CNN on the patient data can be considered as a future work. 
\section{Conclusion}
We proposed a new semi-supervised network for ASL to PET translation. 
The proposed network leverages the residual attention guided mechanism to improve the physiological details. Also, utilizing skip attention gates aids the network to disentangle the ASL and PET relevant features.  
Results showed that incorporating T1w scan as the input and skip attention and residual attention gates improved the results anatomically and physiologically, respectively.

\section*{Acknowledgements}
We are very grateful to the Amsterdam University Medical Center location VUmc for acquiring the PET-data of this study. We especially acknowledge the help of Prof.dr. Ronald Boellaard. This work is financed by the Netherlands Organization
for Scientific Research (NWO), VICI project 016.160.351.
\bibliography{unsrt}
\bibliography{main}
\clearpage


\end{document}